# FpSynt: a fixed-point datapath synthesis tool for embedded systems

Ilya Y. Zhbannikov[1], Gregory W. Donohoe[2]

[1] Institute of Bioinformatics and Computational Biology (IBEST), University of Idaho, Moscow, ID, 83843,
Email: zhba3458@vandals.uidaho.edu

[2] Department of Electrical Engineering, University of Idaho, Moscow, ID, 83843
Email: gdonohoe@uidaho.edu

*Abstract*—**Digital mobile systems must function with low power, small size and weight, and low cost. High-performance desktop microprocessors, with built-in floating point hardware, are not suitable in these cases. For embedded systems, it can be advantageous to implement these calculations with fixed point arithmetic instead. We present an automated fixed-point data path synthesis tool FpSynt for designing embedded applications in fixed-point domain with sufficient accuracy for most applications.**

## I. Introduction

Real-time applications in embedded systems must be implementable in small, lightweight, low-power, low-cost platforms. Often these applications are numerically intensive: real-time speech and video processing, embedded control. A primary attraction of floating points is ease of design – scaling of the data is explicit in the representation and when computational results grow too large, floating point accommodates by reducing the resolution. In embedded applications, these advantages come at a cost: floating point hardware consumes space and power. When emulated in software, floating point may run 20 times slower than implementation in hardware or fixed point. [1], [6]. For these reasons, embedded systems are often built with microprocessors, Digital-Signal Processors (DSP) or Field-Programmable Gate Arrays (FPGA), and hybrid system-on-a-chip (SOC) platforms which do not have hardware floating-point support.

An alternative to floating-point numbers implementation of algorithms is fixed-point arithmetic, which can be implemented with integer-only hardware. The benefits are high speed, low hardware complexity and low cost. When implemented well, fixed-point arithmetic can be accurate enough for most embedded applications.

Designing a fixed-point data path by hand is slow and error-prone. Software design tools can help, but the fixed point tools available nowadays are useful only for simulation, not for synthesis.

We present an automated fixed-point data path synthesis tool FpSynt which offers short design time, overflow control, and minimal arithmetic error. FpSynt synthesizes both VHDL and C code for implementations on FPGAs, microprocessors or Systems-on-Chip. It performs computational error estimation during design process and high-level optimization in order to achieve minimum computational error. We believe that engineering community will benefit from using FpSynt.

## II. Method

### A. Fixed-point design considerations

An overview of FpSynt synthesis pipeline is presented in Figure 1. The algorithm description and input variables specification must be given. The data pre-processing stage is needed to prepare the input data for further analysis. In the next stage, the datapath is synthesized and then the optimization is performed. Finally, the whole fixed-point datapath is synthesized.

The requirements for automated fixed-point datapath design tool are:

- Overflow control
- Maximum numerical accuracy
- Short design time

The challenge is that, in a chain of calculations, the number of bits required to represent a number tends to grow. Much of digital signal processing is based on chains of multiply-add operations. Each multiplication doubles the data width, and each addition has the potential to generate carry bits. In a fixed-width data path, this can lead to overflow, which produces a catastrophic failure. To prevent this, we insert data path formatting operations between numerical operations:

- Before adding two numbers, we pre-scale by dividing both by two (in binary, by shifting right one bit), leaving room on the left for a carry;
- After multiplication, we discard (truncate) half of the product bits to fit the data path; we do pre-scale operations to maximize the number of information-bearing bits retained. Truncation and pre-scaling change the semantics (the scaling of the data), and sacrifice information bits.

The questions are:

- Which formatting operations do we insert in the data path, where do we place them, to minimize the data loss?
- How do we keep track of the scaling so that the mathematical intent of the calculation is preserved, and the original scaling can be restored?

To satisfy those requirements, we introduce a fixed-point notation and a data structure.

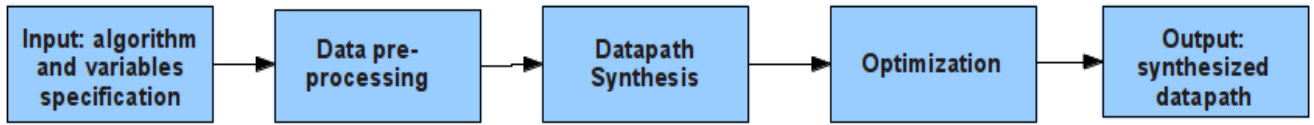

Fig. 1. FpSynt synthesis pipeline.

## B. SIF (Sign-Integer -Fraction) Notation

SIF notation was developed during late 70s in Sandia National Laboratories [3] and adopted for this work. In SIF notation, $S$ is the number of sign bits, $I$ the number of integer bits, and $F$ the number of fraction bits. SIF notation allows to represent any real number within an acceptable range and resolution, capture overflows and track dependencies between variables and scaling factors. In a fixed point binary number (two's complement representation), the bits are divided into bit fields representing sign, integer, and fraction. For example, in the 8-bit two's complement fixed point number 0011.1101, scanning from the left:

- '00' are two sign bits, indicating a positive number
- '11' are integer bits, equivalent to decimal 3
- '.' is the binary point, analogous to a decimal point
- '1101' are fraction bits, equal to decimal 0.8125

Our example has the SIF notation (S/I/F), or (2/2/4). Table I represents examples of this notation.

TABLE I. EXAMPLES OF VALID 16-BIT REPRESENTATION OF BINARY NUMBERS (TWO'S COMPLEMENT).

| SIF-notation | Pattern | Bits |
|---|---|---|
| X(1/0/15) | SFFF FFFF FFFF FFFF | 16 |
| X(2/3/8) | SSII IFFF FFFF F | 13 |
| X(6/3/0) | SSSS SIII | 9 |

## C. Data structure

To cover dependencies between variables and obtain the whole picture of algorithm in global perspective we need to represent the algorithmic structure. A data flow graph (Fig. 2) is chosen as a data structure for fixed-point design tool FPSYNT.

## D. Optimization

On each stage of algorithm, we have many possible choices for shifting and truncating, while preventing overflow and assuring correct data alignment. Once the algorithm is captured in the data flow graph along with the initial data format, we apply a series of optimization steps to minimize the arithmetic error.

Several types of optimization are implemented in FPSYNT tool.

- Combinatorial optimization (Fig. 3) that takes into account all possible fixed-point partitioning of data words; During combinatorial optimization a tree of possible solution is built and FPSYNT takes the solution with minimum error.

- Topological optimization (Fig. 4) which utilizes different possible topologies of the data flow graph: each topology can have different computation error and the goal of this kind of optimization is to find the topology with minimal error;

- Allocating the chains of consecutive additions is a new way of optimization that is based on conception of growing data word size during design process [7].

All of these types of optimization implemented in FpSynt to improve precision and speed-up the design process.

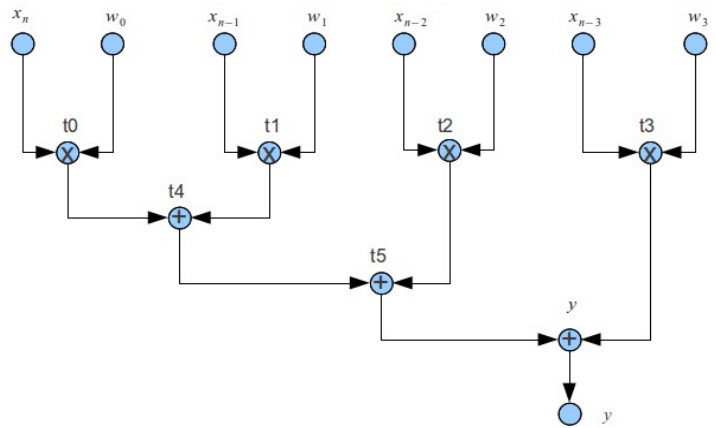

Fig. 2. Data flow graph for 4-tap FIR filter.

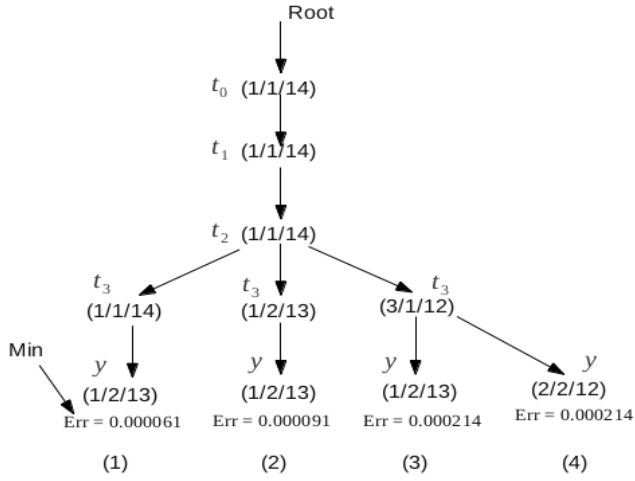

Fig. 3. 3-tap FIR filter data flow graph representations (a) and the combinatorial search tree accompanied to them (b).

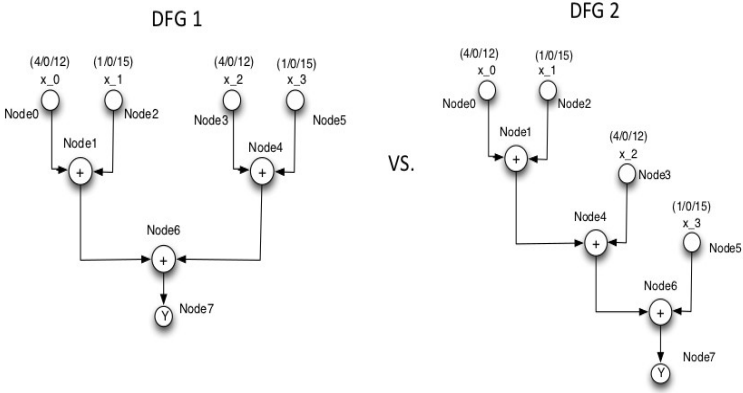

Fig. 4. Two different topologies of the same algorithm.

## III. FPGA IMPLEMENTATION OF FOUR-TAP FIR FILTER WITH FPSYNT

Design the fixed-point datapath is slow and error-prone. We automate the whole datapath design process of 4-tap FIR-filter presented in Fig. 2 with FpSynt. To implement the algorithm on FPGA, assume that input signal is 16 bits wide, and has the following properties:
- One bit for sign;
- Zero bits for integer part;
- Fifteen bits for fraction part,

With coefficients $w_i$ of 0.15, 0.05. 0.45 and 0.35.

FpSynt optimally splits machine word into S-, I-, and F parts minimizing computation error and constructs a fixed-point data path. The summary of program's output is given in Table II.

TABLE II. FPSYNT SUMMARY OUTPUT FOR 4-TAP FIR-FILTER

| Node name | Error | SIF | Operand 1 | Operand 2 |
|---|---|---|---|---|
| t_0 | 0.000031 | (1/1/14) | w_0: (1/0/15) | x_n (7/1/8) |
| t_1 | 0.000031 | (1/1/14) | w_2: (12/0/4) | x_{n-1} (7/1/8) |
| t_2 | 0.000031 | (1/1/14) | w_3: (2/0/14) | x_{n-2} (7/1/8) |
| y | 0.000061 | (1/2/13) | t_3 >> 1* | t_2 >> 1 |

- ">>" represents shift right operation.

To verify the precision of algorithm, the array of 90 input numbers was created, fixed-point algorithm was modeled on FPGA (Xilinx ISE) and results were compared with results from floating-point algorithm. The statistical characteristics of error were computed: max, min, mean and median. Results are given in Table III.

TABLE III. DIFFERENCE BETWEEN FIXED-POINT FPGA IMPLEMENTATION AND FLOATING-POINT ALGORITHM IMPLEMENTATION.

| Type of error | Difference |
|---|---|
| Min | 0 |
| Max | 1.2783e-04 |
| Mean | 5.0621e-05 |
| Median | 5.0613e-05 |

Example of C-code synthesized by FpSynt

```
int X0;
int X1;
int X2;
int X3;
int Y=((((x0*5033165) << 1) +
        (((x1*13421773) >> 2) >> 2) +
        (x2*7549747)) +
        ((x3*11744051) >> 1);
```

A part of VHDL code synthesized by FpSynt:

```
--Coefficients
W0 <= toSIF(0.15);
W1 <= toSIF(0.05);
W2 <= toSIF(0.45);
W3 <= toSIF(0.35);

--Multipliers
T3 <= toSIF(W3*X1);
T2 <= toSIF(W2*X2);
T1 <= toSIF(W1*X3);
T0 <= toSIF(W0*X4);

--Adders
T4 <= toSIF(T0 + T1);
T5 <= toSIF(T4 + T2);
T6 <= toSIF(T5 + T3);
```

## IV. CONCLUSION AND FUTURE WORK

Power, cost, weight and size are major parameters for embedded digital mobile systems, but computations must also be carried out with sufficient accuracy and protection from overflow. Designing an optimal fixed point data path by hand is very difficult; consequently, this powerful technique is under-utilized. We presented an automated approach to generating numerically optimal fixed point data paths. A four-tap FIR filter was designed with FpSynt fixed-point synthesis tool and implemented on FPGA Xilinx Spartan 3E. Experimental results produce an accuracy that is adequate for many applications, in comparison to floating-point model. Future work will be around further optimization of data path synthesis, particularly in exploring pattern optimization [7]. Another approach is to make the synthesis reconfigurable during run-time by running it on System-on-Chip with further uploading to FPGA.

## V. SOFTWARE AVAILABILITY

FpSynt is available under the GNU General Public License from the following GitHub repository:

github.com/izhbannikov/FPSYNT